\documentclass[20pt]{svmult}

\usepackage{url}
\usepackage{hyperref}
\usepackage{mathptmx}       
\usepackage{array}
\usepackage{helvet}         
\usepackage{courier}        
\usepackage{type1cm}        
\usepackage{wrapfig}                            
\usepackage{makeidx}         
\usepackage{graphicx}        
\usepackage{amsmath}                             
\usepackage{multicol}        
\usepackage[bottom]{footmisc}
\usepackage{bm}
\usepackage{setspace}
\usepackage{subcaption}

\usepackage{float}
\usepackage{graphicx,caption}
\usepackage{geometry}
\makeindex             

\begin{document}

\title*{Neutrino oscillations and Leptogenesis}
\author{Anupam Yadav*, Sabeeha Naaz*, Jyotsna Singh* and R.B. Singh*}
\institute{Anupam Yadav*, Sabeeha Naaz*, Dr. Jyotsna Singh* and Dr. R.B. Singh*\at  *Department of Physics, University of Lucknow, Lucknow-226007,India \email{yadavanupam975@gmail.com, sabeehanaaz0786@gmail.com, singh.jyotsnalu@gmail.com, rajendrasinghrb@gmail.com}}
%
%
\maketitle
\vspace{-25mm}
\abstract:{The symmetry breaking of left right symmetric model around few TeV range permits the existence of massive right handed neutrinos or gauge bosons. In this work the decay of lightest right handed neutrinos in a class of minimal left right symmetric model is analysed for the generation of adequate lepton asymmetry. An analytical expression for the lepton asymmetry is developed and the Boltzmann equation are solved. The effect of decay parameter and efficiency factor on the generation of leptogenesis is checked. Lower bound on right handed neutrino mass is imposed. The electroweak sphaleron processes converts the induced lepton asymmetry into baryon asymmetry. In an attempt to achieve the required baryogenesis, we have imposed certain constrains on the parameter space corresponding to low energy neutrino oscillation parameters (especially $\theta_{13}$) and the three phases ( CP, majorana and  higher energy phase).}


\section {Introduction}
Standard model fails to explain the origin of baryonic asymmetry present in the observable universe. The  Big-Bang nucleosynthesis shows that the density of baryons compared to that of photons in the universe is very low; 
$\eta\equiv n_{B} /n_{\gamma} = (2.6-6.3)\times10^{-10}$. 
This ratio $\eta$ can be related to the observed matter-antimatter or baryon-antibaryon asymmetry of the universe as,
$Y_{B}$ $\approx$ $n_B{}/s\approx  \eta / 7.04 = (3.7-8.9)\times10^ {-11}$ \cite{ZZX},
where s denotes the entropy density. In order to produce a net baryon asymmetry in the standard Big-Bang model, three Sakharov necessary conditions must be satisfied :- (a) Baryon number violation, (b) C and CP violation, and (c) The departure from thermal equilibrium \cite{ADS} \cite{EM}. The departure from thermal equilibrium has to be permanent else the baryon asymmetry will be washed out if the thermal equilibrium  is restored. The first two conditions can be investigated only after a particle physics model is specified, whereas the third condition can be discussed in a more general way. 
Among several interesting and viable baryogenesis scenarios proposed in the literature, Fukugita and Yanagida's leptogenesis mechanism has attracted a lot of attention due to the fact that neutrino physics is entering a flourishing era \cite{SSC}\cite{MFTY}.

Further, amongst numerous neutrino mass models under consideration, seesaw mechanism appears as the most attractive model \cite{SFK} for  generation of the tiny neutrino masses reported by the neutrino oscillation experiments \cite{ABM} \cite{NO}. The seesaw mechanism  introduces a massive right handed neutrinos to the model.
Since the right handed neutrinos are very heavy,  usually in neutrino oscillation experiments,  only the left handed neutrino mass or the low energy effective theory is considered for the study of neutrinos \cite{SPAS}. A more careful analysis of seesaw mechanism reveals that if the massive right handed neutrinos are completely ignored in the neutrino analysis or when seesaw mechanism is implemented in context of low energy effective theory, one can miss many essential ingredients of seesaw model required to explain the leptogenesis \cite{JESL}\cite{ASJ}\cite{ASEA}\cite{WR}. In general, without loss of generality one can work with both masses (tiny and massive), in a basis where the charged lepton mass matrix  and right-handed neutrino Majorana mass matrix are diagonal with real eigenvalues \cite{KSB}. In this case,  there will be a total of 18 parameters in neutrino sector and the lepton asymmetry will depend on all of these 18 parameters, where as low energy observables will depend on 9 parameters only. 
In this paper, lepton asymmetry $\eta_{B}$ is estimated from light neutrino mass and mixing parameters by implementing the seesaw mechanism in the context of a class of supersymmetric left$-$right models. By considering a minimal Higgs sector, these models can predict the relation for the Dirac neutrino mass matrix, in a basis where the charged lepton mass matrix is diagonal; where $c \simeq m_{t}/m_{b}$ is determined from the quark sector, and this assumption leaves only the Majorana mass matrix $M_{R}$ to be arbitrary. The three phases of $M_{R}$ can now be removed, leaving a total of 9 parameters which determine both the low energy neutrino masses and mixings as well as the baryon asymmetry. Various attempts have been made by different groups to establish a relationship between leptogenesis and low energy parameters that can be determined by the low energy neutrino experiments.

\vspace{-3mm}
\section{Matter-Antimatter Asymmatry : Numerical Approach:}
\subsection{CP Asymmatry : Conection between Low and High Energy Parameters: }
 In an attempt to explain matter-antimatter asymmetry, leptogenesis is one of the simplest and well motivated mechanism. In this framework new heavy particles are introduced in the theory. The considered framework  must naturally fullfill the three Sakharov conditions.\\
 The new heavy particles considered in Minimal Supersymmetric Model (MSSM) $\cite{KSB}$ are majorana neutrinos $N_{i}$, these particles have a hierarchical mass spectrum   $M_{1}\textless\textless M_{2}\textless M_{3}$, so that the study of evolution of the number density of $N_{1}$ suffices for the study of baryon asymmetry. One important factor that determines the baryon asymmetry, produced by thermal leptogenesis is CP asymmetry (${\epsilon}_{i}$) as discussed above, which can be decomposed as CP asymmetry via $N_{1}$ decays (and similarly $N_{2}$ and $N_{3}$ decays).\\
 
 \begin{equation}   {\epsilon_{1}} = \frac{\Gamma_{N_{1}}\rightarrow lH - {\Gamma_{N_{1}}}\rightarrow \overline{lH}}{\Gamma_{N_{1}}\rightarrow lH + {\Gamma_{N_{1}}}\rightarrow \overline{lH}} \\
 \label{eq-2}
\end{equation}

\begin{equation} \Gamma_{N_{1}} \rightarrow lH = \frac{(m_{D}^{\dagger}m_{D})_{11} M_{1}}{8\pi v^{2}}\\
\label{eq-1}
\end{equation}

Where $m_{D}$ is dirac neutrino mass matrix and $v = 174$
$GeV$ is the VEV (vacuum expectation value) of the Higgs doublet responsible for breaking the electroweak symmetry.\\
For thermal leptogenesis the CP asymmetry produced by decay of $N_{1}$ can be expressed as the sum of a vertex contribution and self energy contribution \cite{SPAS}. In the basis, where right handed neutrino mass matrix is diagonal and real, the CP asymmetry parameter can be written as \cite{KSAB} \cite{MFEA} ,\\

\begin{equation} \bm{\epsilon_{1}} = \bm{\frac{1}{{8\pi v^{2}\boldmath( m_{D}^\dagger}\boldmath m_{D})_{11}}}\sum\limits_{j=2,3}[Im({{m{_D}^\dagger}}{m{_D}})_{1j}^{2}\bm[f_{v}{(\frac{M_{j}}{M_{1}})}^{2}]+\boldmath[f_{s}{(\frac{M_{j}}{M_{1}})}^{2}]]  \\
 \label{eq-3}
\end{equation}

Where $f_{v}(x)$ and $f_{s}(x)$ are the contribution arising from vertex and self energy corrections respectively \cite{MFEA}. For the MSSM case,\\

\hspace{1cm} $f_{v}(x)$ = $\sqrt(x) log (1+\frac{1}{x})$   ;  $f_{s}(x)$ = $\frac{2\sqrt{x}}{x-1}$ \\

\hspace{1cm} $f_{v}(x) + f_{s}(x) $ = $-\frac{3}{2\sqrt{x}}$ ; Here $x = (\frac{M_{j}}{M_{1}})^{2}$\\
 
In a model consisting of three heavy neutrino masses, the CP asymmetry parameter arising from the decay of $N_{1}$ can be written as \cite{MFEA},\\
 
\begin{equation}    \bm{\epsilon_{1}} = \frac{-3}{16\pi v^{2}}{\frac{1}{{\boldmath( m_{D}^\dagger}\boldmath m_{D})_{11}}}[Im{{(m{_D}^\dagger}}{m{_D}})_{12}^{2}\bm{\frac{M_{1}}{M_{2}}}+Im({m{_D}^\dagger}{m{_D}}){_{13}^{2}}\boldmath\frac{M_{1}}{M_{3}}] \\
 \label{eq-4}
\end{equation}

Here $\epsilon_{1}$ depends on the (1,1), (1,2) and (1,3) entries of $(m_{D}^\dagger m_{D})$ .  By the see-saw mechanism light neutrino mass matrix can be connected to heavy neutrino mass matrix by the expression \cite{PMTY},\\

\begin{equation}  m_{\nu} = -m_{D} M_{N}^{-1} m_{D}^{T}\\
 \label{eq-5}
\end{equation}

where $m_{D}$ is,

\begin{equation} m_{D}=c m_{l} = c* diag (m_{e},m_{\mu},m_{\tau}) \\
 \label{eq-6}
\end{equation}

here $m_{l}$ is charged lepton mass matrix and c is defined as $c=m_{t}/m_{b}$.\\

 In the case of Neutrino, flavor eigen states $(\nu_{e},\nu_{\mu},\nu_{\tau})$ and mass eigen states $(\nu_{1},\nu_{2},\nu_{3})$ can be connected as \cite{ZMMN} ,\\
\begin{equation}  m_{\nu} = U m_{\nu}^{diag} U^{\dagger}\\
 \label{eq-7}
\end{equation}

Where $m_{\nu}^{diag} = diag (m_{1},m_{2},m_{3})$ and U is a $3\times3$ mixing matrix consisting of, majorana phases and dirac phase \cite{SMB}. The solution for majorana mass matrix can be expressed as,\\
\begin{equation}  M_{N} = c^{2} m_{l} m_{\nu}^{-1} m_{l}^{T}\\
 \label{8}
\end{equation}

 \begin{equation} = \frac{c^{2}m_{\tau}^{2}}{m_{1}}  \begin{bmatrix}
\frac{ m_{e}}{m_{\tau}}      & 0     & 0 \\
0                            & \frac{m_{\mu}}{m_{\tau}} & 0 \\
0                            & 0                       & 1
  
 \end{bmatrix}  U_{PMNS}P^{2} \begin{bmatrix}
1      & 0     & 0 \\
0                            & \frac{m_{1}}{m_{2}} & 0 \\
0                            & 0                       & \frac{m_{1}}{m_{3}}
  
 \end{bmatrix}  U_{PMNS}^{T}  \begin{bmatrix}
\frac{ m_{e}}{m_{\tau}}      & 0     & 0 \\
0                            & \frac{m_{\mu}}{m_{\tau}} & 0 \\
0                            & 0                       & 1
  
 \end{bmatrix} \\
 \label{eq-9}
 \end{equation}\\
 
 The matrix P contains two majorana phases $P = (e^{i\alpha}, e^{i\beta}, 1)$.
 In an attempt to get proper form of $M_{N}$, the values of different neutrino parameters are expressed in  a small expansion parameter \cite{KSAB} \cite{GCB} \cite{WBMP} $\gamma$, which is defined as,\\

\hspace{3cm}    $\gamma$ = $\frac {\bm{m_{\mu}}}{\bm{m_{\tau}}}$ $\simeq 0.059$ \\

 In terms of small expansion parameters $\gamma$, different parameters can be expressed as ,\\

 $m_{e}$ = $a\gamma^{3}m_{\tau}$, ${\frac{\bm{m_{1}}}{\bm{m_{3}}}}$ = $a_{13}\gamma$, $\frac{\bm{m_{1}}}{\bm{m_{2}}}$ = $\tan^{2}{\theta_{12}}$ + $a_{12}\gamma$ , $\theta_{23}$  =  $\boldmath\frac{\pi}{4}$ + $t_{23}\gamma$, $\beta$ = $\alpha+\pi/2+b\gamma$ \\
 
 and  $\sin\theta_{13}$ = $\theta_{13}$, since $\theta_{13}$ is very small. \\
 
Where $a$, $a_{13}$, $t_{13}$ , $t_{23}$ are parameters of $O(1)$ , with a = 1.400, $a_{13}$= 1.3, $a_{12}$ = 1 and $b= 1$.

Now the right handed mass matrix can be written as,\\
 
 \begin{equation} M_{N}= \begin{bmatrix}
   P\gamma^{5}       &Q\gamma^{3}       &R \gamma^{2}\\
   S\gamma^{3}           &T\gamma^{2}    &U\gamma\\
   V\gamma^{2}           &W\gamma        &X
  \end{bmatrix}
  \end{equation}\\
  
  The unitary matrices of $M_{N}$ can be difined as $(KU_{3}U_{2}U_{1})$;\\

 \begin{equation} (KU_{3}U_{2}U_{1})M_{N}(KU_{3}U_{2}U_{1})^{T} =   \begin{bmatrix}
   \mid M_{1} \mid       &0        &0\\
   0           &\mid M_{2} \mid    &0\\
   0           &0        & \mid M_{3} \mid
  \end{bmatrix}
  \end{equation}\\

Where $K= diag(k_{1},k_{2},k_{3})$ with $k_{1}= e^{-i\phi_{1}/2}$, where $\phi_{1},\phi_{2}, \phi_{3}$ are phase factors which make each right handed neutrino masses real.\\
 
  The values of masses $m_{1}$, $m_{2}$, and $m_{3}$ considered in this work are ,\\

 \hspace{2cm} $m_{1} = 0.0027\times10^{-9} GeV $\\

 \hspace{2cm} $m_{2} = 0.0068\times10^{-9} GeV $\\

 \hspace{2cm} $m_{3} = 0.0380\times10^{-9} GeV $\\

 After the expansion in terms of small parameter, the values of $(m_{D}^{\dagger}m_{D})_{11}$ , $(m_{D}^{\dagger}m_{D})_{12}^{2}$  and $(m_{D}^{\dagger}m_{D})_{13}^{2}$,  which are connected to matrix $M_{N}$ takes the form,\\

 \begin{equation}(m_{D}^{\dagger}m_{D})_{11} = \frac{8a^{2}c^{2}m_{\tau}^{2}\gamma^{2}\cos^{2}\theta_{12}\sin^{2}\theta_{12} \theta_{13}^{2}\tan^{2}\theta_{12}}
 {{a_{13}\cos^{4}\theta_{12}[4a_{13}(a_{12}^{2}-b^{2})\cos2\theta_{12})+a_{13}(a_{12}^{2}+4b^{2})(3+\cos4\theta_{12})}]}
 \end{equation}\\

 \begin{equation}(m_{D}^{\dagger}m_{D})_{12}^{2} =  \frac{2a^{2}c^{4}m_{\tau}^{4}\gamma^{2}\tan^{2}\theta_{12}e^{-i(\phi_{1}-\phi_{2})}e^{-2i(2\alpha+\delta)} [{-4\theta_{13}^{2}\cos2\theta_{12}+4\theta_{13}^{2}}]^{2}}
{{[3a_{12}a_{13}-2ia_{13}b+4\cos^{2}\theta_{12}a_{12}a_{13}+a_{13}(a_{12}+2ib)\cos4\theta_{12}]}^{2}}\end{equation}\\

 \begin{equation}(m_{D}^{\dagger}m_{D})_{13}^{2} = \frac{2a^{2}c^{4}m_{\tau}^{4}\sin^{2}\theta_{12}e^{-i(\phi_{1}-\phi_{3})}[a_{13}^{2}\cos^{2}\theta_{12}\gamma^{6}+e^{-2i(2\alpha+\delta)} \theta_{13}^{2}\sin^{2}\theta_{12}\gamma^{4}+2a_{13}\cos\theta_{12}e^{-i(2\alpha+\delta)}\theta_{13}\sin\theta_{12}\gamma^{5}]}{[{a_{13}\gamma^{4}\cos^{4}\theta_{12}(a_{12}-2ib+(a_{12}+2ib)\cos2\theta_{12})}]^{2}}\end{equation}\\

Now the Eq. (4),  representing the CP asymmetry can be expressed as,\\

\begin{equation}    \bm{\epsilon_{1}} = {\frac{-3}{16\pi v^{2}}\frac{1}{{\boldmath( M_{D}^\dagger}\boldmath M_{D})_{11}}}[{2ABC+D(A^{2}-B^{2})/(A^{2}+B^{2})^{2}\times(\bm{\frac{M_{1}}{M_{2}}})}+2EFG+H(E^{2}-F^{2})/(E^{2}+F^{2})^{2}\times(\boldmath\frac{M_{1}}{M_{3}})] \\
 \label{eq-15}
\end{equation}

Different terms used in the above expression are,\\

\hspace{1cm}$A$ = $(3a_{12}a_{13}+4\cos^{2}\theta_{12}a_{12}a_{13}+a_{12}a_{13}\cos4\theta_{12})$\\

\hspace{1cm}$B = i(2ba_{13}\cos4\theta_{12}-2a_{13}b)$\\

\hspace{1cm}$C = 2a^{2}c^{4}m_{\tau}^{4}\gamma^{2}\tan^{2}\theta_{12}[\cos(\phi_{1}-\phi_{2})\cos(4\alpha+2\delta)-$\\

\hspace{1cm}$\sin(\phi_{1}-\phi_{2})\sin(4\alpha+2\delta)][{-4\theta_{13}^{2}\cos2\theta_{12}+4\theta_{13}^{2}}]^{2} $\\

\hspace{1cm}$D = 2a^{2}c^{4}m_{\tau}^{4}\gamma^{2}\tan^{2}\theta_{12}[\sin(\phi_{1}-\phi_{2})\cos(4\alpha+2\delta)+$\\

\hspace{1cm}$\cos(\phi_{1}-\phi_{2})\sin(4\alpha+2\delta)][{-4\theta_{13}^{2}\cos2\theta_{12}+4\theta_{13}^{2}}]^{2} $\\

\hspace{1cm}$E = {2\sqrt(a_{13})\gamma^{2}cos^{2}\theta_{12}b(\cos2\theta_{12}-1)}$\\

\hspace{1cm}$F = i[{\sqrt(a_{13})\gamma^{2}cos^{2}\theta_{12}a_{12}(1+\cos2\theta_{12})}]$\\

\hspace{1cm}$G = 2a^{2}c^{4}m_{\tau}^{4}\sin^{2}\theta_{12}\cos(\phi_{1}-\phi_{3}) $\\

\hspace{1cm}$[{a_{13}^{2}\cos^{2}\theta_{12}\gamma^{6}+\theta^{2}_{13}\sin^{2}\theta_{12}\gamma^{4}\cos(4\alpha+2\delta)+2a_{13}\cos\theta_{12}\theta_{13}\sin\theta_{12}\gamma^{5}\cos(2\alpha+\delta)}]$\\

\hspace{1cm}$-2a^{2}c^{4}m_{\tau}^{4}\sin^{2}\theta_{12}\sin(\phi_{1}-\phi_{3})$\\

\hspace{1cm}$[\theta^{2}_{13}\sin^{2}\theta_{12}\gamma^{4}\sin(4\alpha+2\delta)+2a_{13}\cos\theta_{12}\theta_{13}\sin\theta_{12}\gamma^{5}\sin(2\alpha+\delta)]$\\

\hspace{1cm}$H = 2a^{2}c^{4}m_{\tau}^{4}\sin^{2}\theta_{12}\sin(\phi_{1}-\phi_{3}) $\\

\hspace{1cm}$[{a_{13}^{2}\cos^{2}\theta_{12}\gamma^{6}+\theta^{2}_{13}\sin^{2}\theta_{12}\gamma^{4}\cos(4\alpha+2\delta)+2a_{13}\cos\theta_{12}\theta_{13}\sin\theta_{12}\gamma^{5}\cos(2\alpha+\delta)}]$\\

\hspace{1cm}$+2a^{2}c^{4}m_{\tau}^{4}\sin^{2}\theta_{12}\cos(\phi_{1}-\phi_{3})$\\

\hspace{1cm}$[\theta^{2}_{13}\sin^{2}\theta_{12}\gamma^{4}\sin(4\alpha+2\delta)+2a_{13}\cos\theta_{12}\theta_{13}\sin\theta_{12}\gamma^{5}\sin(2\alpha+\delta)]$\\

The hightest contributing factor in the CP asymmetry  $\epsilon_{1}$ can be stated as,\\

\begin{equation}
 \bm{\epsilon_{1}}\propto\theta_{13}^{2}\cos(\phi_{1}-\phi_{2})\cos2(2\alpha+\delta)-\theta_{13}^{2}\sin(\phi_{1}-\phi_{2})\sin2(2\alpha+\delta)+[\cos(\phi_{1}-\phi_{3})+\sin(\phi_{1}-\phi_{3})]/\theta_{13}^{2}+ .......
\end{equation}\\
The Davidson-Ibarra (DI) bound on a CP asymmetry is given as,
\begin{equation}
 |\epsilon_{1}(M_{1}^{d=5}, \Bar m)| \leq |\epsilon_{1}^{max, d=5}(M_{1}, \Bar m)| = \frac{3}{16\pi}\frac{M_{1}}{v^{2}}(m_{3} - m_{1})
\end{equation}

\subsection{Quntitative Calculation of the Abundance of Right handed Neutrino:}

With in the minimal framework where we assume that the initial temprature $T_{i}$ is larger than $M_{1}$, is the mass of the lightest heavy neutrino $N_{1}$ we neglect the decays of two heavier neutrinos $N_{2}$ and $N_{3}$ assuming that it does not influence the final value of B-L asymmetry. Let us consider $N_{N_{1}}$ indicates the right handed neutrino abundance and $N^{eq}_{N_{1}}$ indicates the thermal equilibrium values of right handed neutrino. The Boltzmann equation for $N_{N_{1}}$ can be given as

\begin{equation}
 \frac{dN_{N_{1}}}{dz} = -(D + S)(N_{N_{1}} - N^{eq}_{N_{1}})
\end{equation}

From the above equation where $z = M_{1}/T$ we notice that the lepton asymmetry will be generated when lightest right handed neutrino is 'out-of-thermal equilibrium' ($N_{N_{1}}$ $\neq$ $N^{eq}_{N_{1}}$). As $N_{N_{1}}^{eq}$ drpos with universe temprature T, the out-of-thermal equilibrium can be satisfied, when the universe is cooling down. If the Hubble expansion rate is denoted by H then,
D = $\frac{\Gamma_{D}}{H z}$ accounts for decays and inverse decays where as 
S = $\frac{\Gamma_{S}}{H z}$ represents $\Delta L = 1$ scattering term. The Hubble expansion rate is given by,\\

\hspace{1cm} H $\simeq \sqrt{\frac{8\pi^{3}g_{*}}{90}}\frac{M_{1}^{2}}{M_{PI}}\frac{1}{z^{2}}$\\

where, $g_{*} = g_{SM} = 106.75$ is the total number of degrees of freedom, and $M_{PI} = 1.22\times 10^{19}$ $GeV$, is the Planck mass.\\

The two terms D and S  depend on effective neutrino mass ($\tilde m_{1}$), which is expressed as,\\
 
  \begin{equation}\tilde m_{1} = \frac {(m_{D}^{\dagger}m_{D})_{11}}{M_{1}}\end{equation}\\
  
  Effective neutrino mass has to be compared with the equilibrium neutrino mass ($m_{*}$), which is expressed as,\\
  
  \begin{equation} m_{*} = \frac{16\pi^{5/2}\sqrt{g_{*}}v^{2}}{3\sqrt{5}M_{pl}} \simeq 1.08\times10^{-3} eV
  \end{equation} \\
  
  The deacy parameter
   \begin{equation}
    K = \frac{\Gamma_{D}(z = \infty )}{H (z = 1)} = \frac{\tilde m_{1}}{m_{*}}
   \end{equation}
   
   \begin{figure}[H]
    \includegraphics[scale=.70]{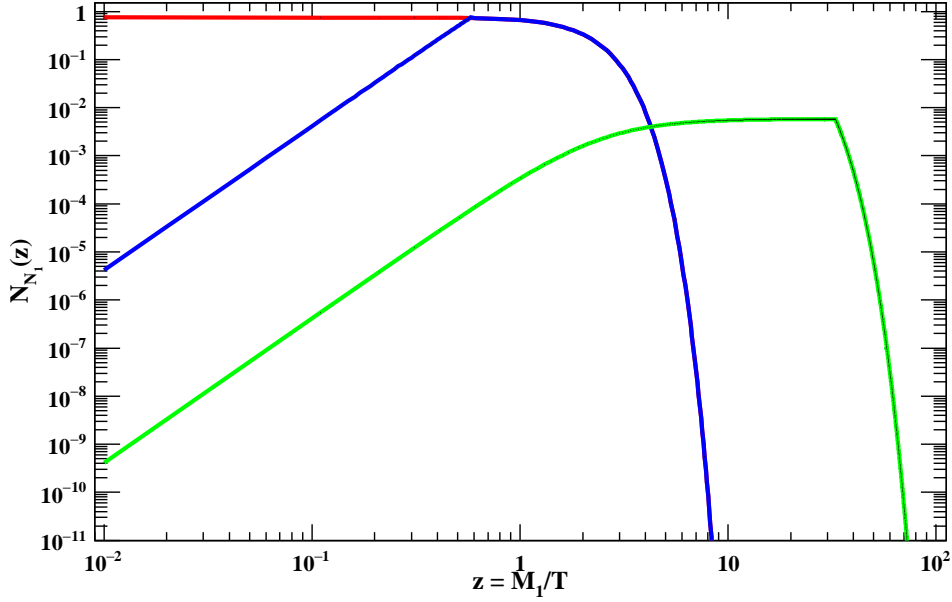}
    \caption{Result of heavy neutrino production in the case of zero initial abundance at $K = 10^{-2}$ (green line), $K = 100$ (blue line) and thermal initial abundance (red line) at $z_{eq}$.}
    \label{Fig:1}
   \end{figure}

   For the calculation of the abundance of right handed neutrino the decay parameter (K) plays a key role. For $K\textgreater\textgreater1$, the life time of right handed neutrino is much shorter than the age of the universe, $t = H^{-1}/2 $ $(z = 1)$ and the right handed neutrino decays and inverse decays many times before they become non-relativistic. In this case the abundance of right handed neutrino resembles very closely to the equilibrium distribution as shown in Fig:1. On the other hand when $K\textless\textless1$, when right handed neutrino are already fully non-relativistic and the bulk of right handed neutrinos decay completely out of equilibrium hence in this case their equilibrium abundance is exponentially supressed by Boltzmann factor. This can be checked in Fig:1. Where we can see that the pattern for the production of the right handed neutrino at $K = 100$ overlaps the pattern produced for the same at $z_{eq}$, after $z = 1$. Where as the production pattern of right handed neutrino for $K \textless 1$ with stands for larger value of $z$.
\subsection{Calculation of Baryon, Lepton Asymmetry and Efficiency factor:}
The Boltzmann equation for $N_{B-L}$ can be given as,

\begin{equation}
 \frac{dN_{B-L}}{dz} = -\epsilon_{1} D (N_{N_{1}} - N^{eq}_{N_{1}}) -  W N_{B-L} 
\end{equation}

where, $ W = \frac{\Gamma_{w}}{H z}$ contributes two washout term, $ W = W_{0} + \Delta W$; the first term depends only on $\tilde m_{1}$, while the second term depends on the product $M_{1}\Bar m^{2}$, where $\Bar m^{2} = m_{1}^{2} + m_{2}^{2} + m_{3}^{2}$ is sum of the light neutrino masses squared.  \\
The washout term is the term that tends to re-equilibrate the number of leptons and antileptons destroying the asymmetry generated by the CP violating term. It is simply a statistical re-equilibrating term that has to be present in order to respect the Sakharov third condition. \\
Solution of above Boltzmann equation for $N_{B-L}$ is the sum of two term,
\begin{equation}
 N_{B-L}(z) = N_{B-L}^{i}e^{-\int_{z_{i}}^{z} dz' W(z')} - \frac{3}{4}\epsilon_{1}k(z; \tilde m_{1}, M_{1}\Bar m^{2})
\end{equation}

First term of the above equation accounts for possible generation of $B-L$ asymmetry before $N_{1}$ decays
and the second part of the above equation expressed in terms of the efficiency factor ($k$) and CP asymmetry ($\epsilon_{1}$), describes $B-L$ generation from $N_{1}$ decays. In our analysis we have neglected the contribution arising from the first term. The efficiency factor does not depend on CP asymmetry $\epsilon_{1}$.\\
A global expression for the efficiency factor can be expressed as a sum of positive contribution $k_{f}^{+}(K)$ when $K\textgreater\textgreater1$ and negative contribution $k_{f}^{-}(K)$ when $K\textless\textless1$.


\begin{center}
 $k_{f}(K) = k_{f}^{+}(K) + k_{f}^{-}(K)$
\end{center}

 \begin{equation}
  k_{f}^{-} = -2 e^{-\frac{2}{3}(N(K) + \frac{3}{4}K\alpha_{s})} (e^{\frac{2}{3}\tilde N(K)} - 1) 
 \end{equation}

\begin{equation}
 k_{f}^{+} = \frac{2}{z_{B}(K)Kj(z_{B})^{2}}(1 - e^{-\frac{2}{3}z_{B}(K)Kj(z_{B})^{2}N_{N_{1}}(z_{eq})j(z_{eq})^{-1} })
\end{equation}

Here, $z_{B} = \frac{M_{1}}{T_{B}}$, where $T_{B}$ is baryogenesis temperature and $z_{B}(K)$ is defined as,\\

\begin{equation}
  z_{B}(K) \simeq 1 + \frac{1}{2}\ln(1 + \frac{\pi K^{2}}{1024}[\ln(\frac{3125\pi K^{2}}{1024})]^{5})
\end{equation}
\begin{equation}j(z) = \frac{D + S}{D} \approx [\frac{1}{a}\ln(1 + \frac{a}{z}) + \frac{K_{s}}{K z}] (1 + \frac{15}{8z})
\end{equation}

\begin{equation}
 a = \frac{K}{K_{s}\ln(M_{1}/M_{h})} = \frac{8\pi^{2}}{9\ln(M_{1}/M_{h})} 
\end{equation}
Scattering parameter,
\begin{equation}
K{s} = \frac{\tilde m_{1}}{m_{*}^{s}}
\end{equation}

\begin{equation}
 m_{*}^{s} = \frac{4\pi^{2}}{9}\frac{g_{N_{1}}v^{2}}{m_{t}^{2}}m_{*} \simeq 10 m_{*}
 \end{equation}
\begin{equation}
 \alpha_{s} = \frac{2K_{s}}{3K} + \frac{15}{8}
\end{equation}
\begin{equation}
 \tilde N(K) = \frac{2N(K)z_{eq}^{3}}{((9\pi)^{c} + (2z_{eq}^{3})^{c})^{1/c}}
 \end{equation}
 \hspace{1cm} $z_{eq} = (\frac{6}{K})^{1/3};  c = 0.7;        N(K) = 9\pi K/16 $ \\
 
 In this efficiency factor both decay and scattering are considered. The case without scattering can be recovered by substituting $\alpha_{s} = 0$ in $k_{f}^{-}$ and $j = 1$ in $k_{f}^{+}$. Fig:2 illustrates the $N_{B-L}$ asymmetry produced for two different values of $K$ ($K\textless\textless1$ and $K\textgreater\textgreater1$). \\
 Electroweak sphaleron processes is responsible for the conversion of lepton asymmetry into baryon asymmetry.
 The baryon asymmetry $n_{B}$ produced through the sphaleron transition of lepton asymmetry $Y_{L}$,  while the quantum number B-L remains conserve, is given by,\\

 \begin{equation}
 Y_{B} = \frac {n_{B}}{s} = C Y_{B-L} = CY_{L}\end{equation}\\
 
 Where $C = \frac {8N_{f} + 4N_{H}}{22N_{f} + 13N_{H}}$; $N_{f}$ is the number of fermionic family, $N_{H}$ is the number of Higgs doublets and $s=7.04$ $n_{\gamma}$.\\
 \begin{equation}Y_{L} = \frac{(n_{L}-\overline n_{L})}{s} = {\sum}_{i=1}^{3} \frac {\epsilon_{i}k_{i}}{g_{*i}}\end{equation}\\
After substituting the eq(34) in eq(33), we get,
\begin{equation}Y_{B} = C \frac{\epsilon_{1}k_{1}}{g_{*1}}\end{equation}\\
In our work we have to used $k_{f}$ in place of $k_{1}$ and $g_{*}$ in place of $g_{*1}$.

\begin{figure}[H]
    \includegraphics[scale=.70]{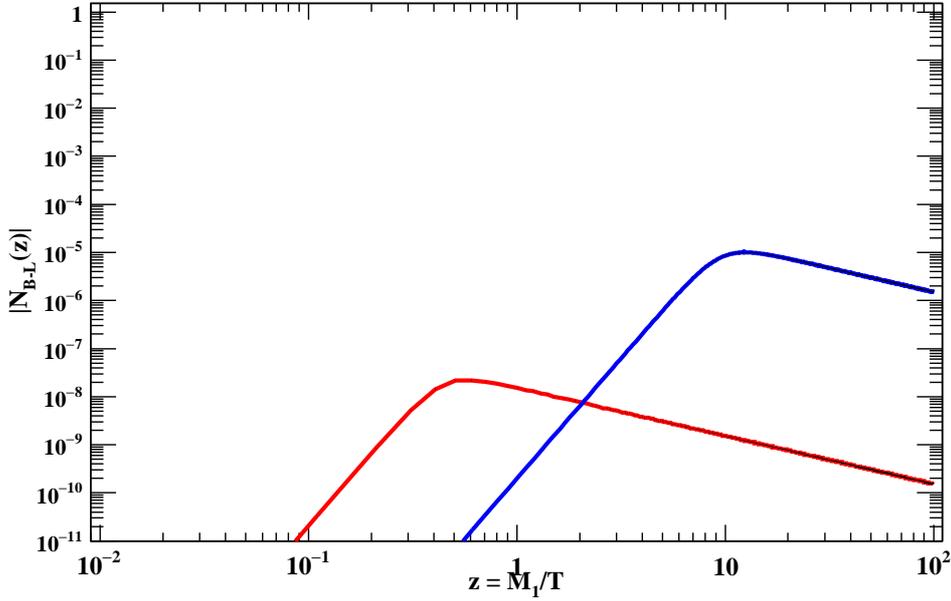}
    \caption{Result of baryon and lepton asymmetry in the case of thermal initial abundance at $K = 10^{-2}$ (blue line) and $K = 100$ (red line).}
    \label{Fig:2}
   \end{figure}
If we assume that the universe reheats to a thermal bath composed of particles with case interaction after inflation, the final baryon to photon number ratio $\eta_{B}$ can be estimated as the product of three suppresion factor \\
1) Leptonic CP asymmetry $\epsilon_{i}$.\\
2) An efficiency factor $k_{f}$ ,arising due to washout processes and scattering, decays and inverse decays.\\
3) A reduction factor due to chemical equilibrium charge conservation and the redistribution of asymmetry among different particle species.\\

\section{Bounds on the Mass of Right handed Neutrino:}
The light neutrino masses can be either quasi degenerate or hierarchical, with $m_{2} - m_{1}\textless\textless m_{3} - m_{2}$ as normal hierarchy and $m_{2} - m_{1}\textgreater\textgreater m_{3} - m_{2}$ as inverted hierarchy. The maximal CP asymmetry $\epsilon_{1}^{max}$ depends on $M_{1}$ and $\tilde m_{1}$ and via the light neutrino masses $m_{i}$, on absolute neutrino mass scale $\Bar m$. For given light neutrino masses, $\epsilon_{1}$ is maximized in the limit $\frac{m_{1}}{\tilde m_{1}}$ tending to zero. The upper bound on CP asymmetry $\epsilon_{1}$ as a function of $M_{1}$ and $\tilde m_{1}$ is expressed in eq(17). Eq(17) reaches its maximum value for fully hierarchical neutrinos with $m_{1} = 0$ and $m_{3} = m_{atm} = \sqrt{\Delta m_{atm}^{2}}$ and can be written as,\\
\begin{equation}
 \epsilon_{1}^{max}(M_{1}, \Bar m) = \frac{3}{16\pi}\frac{M_{1}}{v^{2}} m_{3}
\end{equation}
From the latest neutrino oscillation experimental results the values of $\Delta m_{atm}^{2}$ and $\Delta m_{sol}^{2}$ are as follows \cite{PF}, \\
\begin{center}
 $\Delta m_{atm}^{2} = (2.5 \pm 0.03) \times 10^{-3} eV^{2}$ (NH)
\end{center}
\begin{center}
 $\Delta m_{atm}^{2} = 2.45^{+0.03}_{-0.04} \times 10^{-3} eV^{2}$ (IH)
\end{center}
\begin{center}
 $\Delta m_{sol}^{2} = 7.55^{+0.2}_{-0.16} \times 10^{-5} eV^{2}$
\end{center}
\begin{center}
 $m_{atm} = (0.05 \pm 0.003 ) eV $
\end{center}
A recent combined analysis of baryon to photon ratio \cite{CMB} is,
\begin{equation}
 \eta^{CMB}_{B} = 6.0^{+0.8}_{-1.1} \times 10^{-10}
\end{equation}
The CP asymmetry in terms of $\eta_{B}^{CMB}$ can be written as, \cite{W},\\
\begin{equation}
 \epsilon^{CMB}_{1} \simeq 6.3\times 10^{-8} (\frac{\eta^{CMB}_{B}}{6\times 10^{-10}}) k_{f}^{-1}
\end{equation}
For maximal baryon asymmetry $\eta_{B}^{max}$ we get maximal CP asymmetry, which is evident from eq(38). By CMB constraint we get $\eta^{max}_{B}\geq \eta^{CMB}_{B}$.
Since $m_{atm}$ is fixed quantity then from eq(36), maximal value of CP asymmetry will depends only on $M_{1}$,

\begin{equation}
 \epsilon_{1}^{max}(M_{1}) = \frac{3}{16\pi}\frac{M_{1}}{v^{2}} m_{3} \simeq 10^{-6} (\frac{M_{1}}{10^{10} GeV}) (\frac{m_{atm}}{0.05 eV})
\end{equation}


\begin{equation}
 M_{1}\textgreater M_{1}^{min} \simeq 6.4\times 10^{8} GeV (\frac{\eta^{CMB}_{B}}{6\times 10^{-10}}) (\frac{0.05 eV}{m_{atm}}) k_{f}^{-1}
\end{equation}

Bounds on $M_{1}$ depends on the combination $\eta^{CMB}_{B}/m_{atm}$,\\

\begin{equation}
 M_{1}^{min} (\tilde m_{1}) = (6.4 \pm 0.6)\times10^{8} GeV k_{f}^{-1} \geq 4\times 10^{8} k_{f}^{-1} (\tilde m_{1})
\end{equation}

\begin{figure}[H]
    \includegraphics[scale=.70]{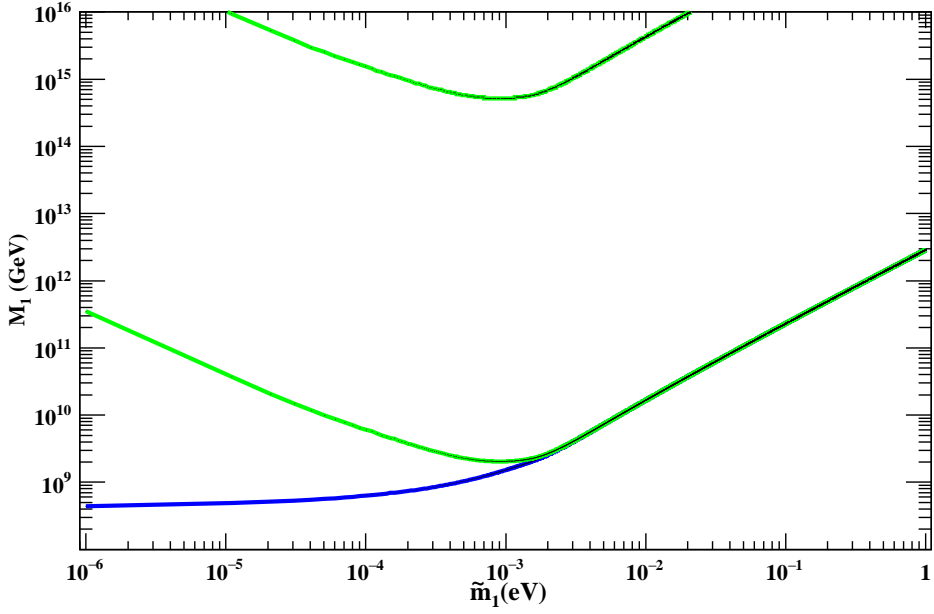}
    \caption{Lower bounds on $M_{1}$ for thermal initial abundance (blue line) and for zero initial abundance (green line).}
    \label{Fig:3}
   \end{figure}
The observed baryon asymmetry $\eta_{B} \approx 10^{-10}$ sets a lower bound limit on $\epsilon_{1}$ and therefore on $M_{1}$. If right handed neutrinos are produced thermally then $\eta_{B} \leq 10^{-2} \epsilon_{1}$ and $M_{1} \textgreater 10^{8}$ GeV. From Fig:3 the lower bound on right handed neutrino mass $M_{1}$ in thermal initial abundance case is $\sim 5 \times 10^{8}$ GeV and in zero initial abundance case is $\sim 2 \times 10^{9}$ GeV.

 \section{Results and Discussion:} 
  
In developing  the  above  formulation,  we  have  assumed  low  energy  supersymmetry, where the Dirac neutrino mass  matrix  has a determined structure. As a result, we have connected the lepton  asymmetry with measurable low energy  neutrino  parameters. Here right  handed  neutrino  masses  are  not  independent  of  CP  asymmetry  parameter. Our study is restricted to the case where the baryon asymmetry is generated only  due  to  the decay of right handed neutrinos. Three right handed neutrinos, having hierarchical mass structure is considered in this work. 
In the model considered for the generation of baryon asymmetry it is assumed that in early universe, at temperature of order $N_{1}$, the main thermal process, which entered  in the production  of  lepton  asymmetry was  the  decay of lightest  right  handed  neutrino.
In  order  to estimate the  baryon  asymmetry  arising due  to the formulated analytical  expression,  the  dilution  factor, often referred  as the  efficiency  factor $k_{f}$, that takes into account the washout processes (inverse decays and lepton number violating scattering) has to be known a priori.
Hence the solution of Boltzmann equation for the abundance of right handed neutrino and for the generation of $N_{B - L}$ is performed. While performing the solution of Boltzmann equation the effect of various parameters i.e. washout effect ($W$), decay parameter ($K$), efficiency ($k_{f}$) and the ratio of right handed neutrino mass to the temprature ($z = M_{1}/T$) are studied  for the generation of excess $N_{1}$ and $N_{B-L}$, to achieve observed value of baryogenesis.The generated lepton asymmetry gets converted to  the  baryon   asymmetry in the  presence of the  sphaleron induced  anomalous B-L violating processes before the electroweak  phase  transition.
The effect of the decay parameter $K$, $z$ and efficiency factor $k_{f}$ on the value of $N_{N_{1}}$ and $N_{B-L}$ is shown in Fig:1 and Fig:2. The efficiency factor considered in our work takes into account scattering and decay both.
The lowest bound on the right handed neutrino is also examined and is illustrated in Fig:3. The lowest bound achieved in this work is $M_{1} \geq 5 \times 10^{8}$.

\begin{figure}[H]
\includegraphics[scale=.70]{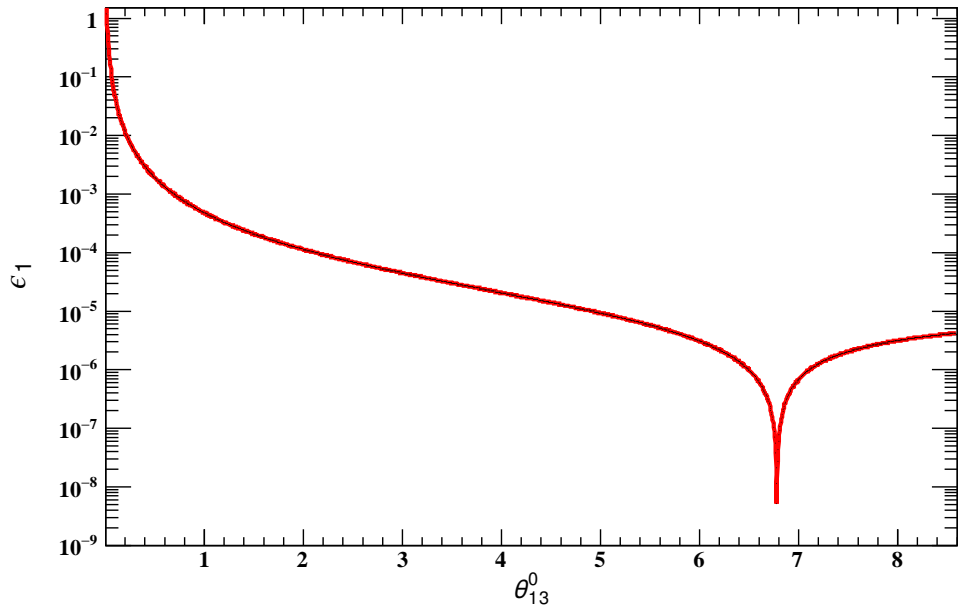}
\caption{ Evolution of CP asymmetry parameter $\epsilon_{1}$ using analytical results as a function of neutrino oscillation angle $\theta_{13}$.}
\label{fig:1}       
\end{figure}


\begin{figure}[H]
\includegraphics[scale=.70]{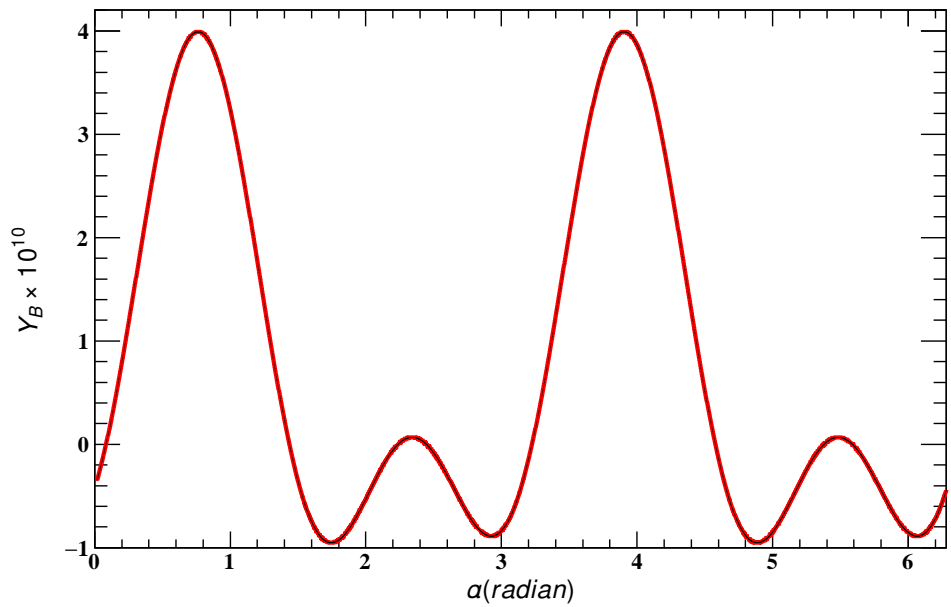}
\caption{Plot for Baryon asymmetry parameter $Y_{B}$ as a function of majorana phase angle $\alpha$.}
\label{fig:2}       
\end{figure}

\begin{figure}[H]
\includegraphics[scale=.70]{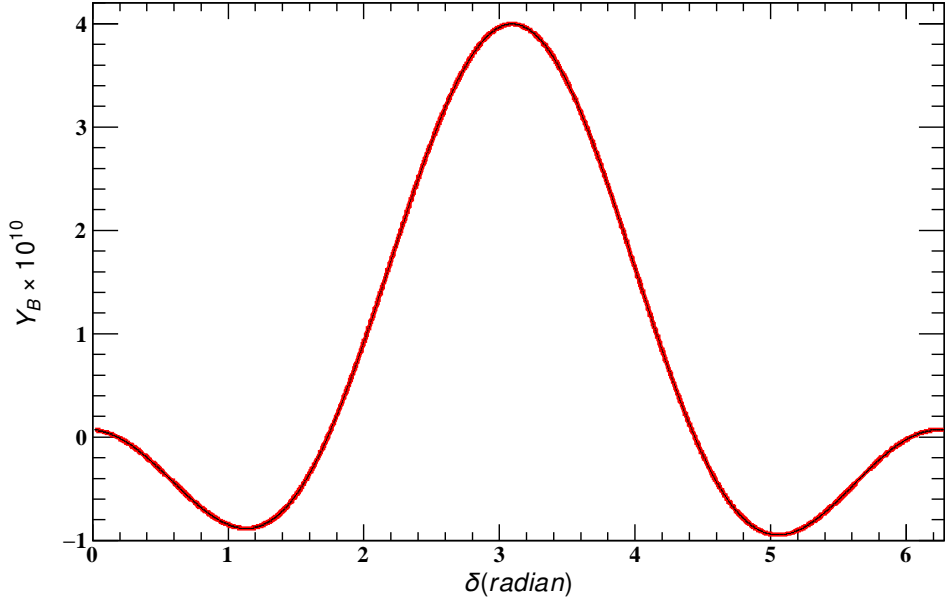}
\caption{Evolution of Baryon asymmetry parameter $Y_{B}$ as a function of dirac phase angle $\delta$.}
\label{fig:3}      
\end{figure}

The  parameter  space  corresponding  to  the  parameters $\theta_{13}$, the CP  phase $\delta$ and  the Majorana phase $\alpha$ are scanned. The value of $\epsilon_{1}$  is checked for a given set of parameters and  the current best fit value of $\theta_{13}$ \cite{PFD} as expressed in table 1. In our analysis if $\epsilon_{1} \textless 1.3\times 10^{-7}$ \cite{KSAB} , the induced baryon asymmetry  would  be  too small  to explain the experimental observations. In Fig.4 we observe that at current best  fit value of oscillation  angle $\theta_{13}$, the value of $\epsilon_{1}$ is sufficient  enough  to generate  observable the baryogengesis  signals  at  low  energy  neutrino  experiments. The  value  of  leptogenesis change  with  the  variation  in  input  parameters.\\

 \begin{table}[] 
  \caption{The value of  various selected parameters used for our analysis.}
  \centering
  \begin{tabular}{|c| c| c| c| c| c| c| c| c| c| c| c|}
   \hline
   a & b & c & $a_{12}$ & $a_{13}$ & $\gamma$ & $M_{1}$ (GeV)& $M_{2}$ (GeV) & $M_{3}$ (GeV) &$m_{\tau}$ (GeV) & $\theta_{12}$ & $\theta_{13}$\\[1.0ex]
   \hline
   1.4 & 1.0 & 41.1  & 1.0 & 1.3 & 0.059 & 1$\times10^{9}$ & 8.7$\times10^{11}$ & 2.6$\times10^{14}$ & 1.77 & 34.5$^{0}$ & 8.45$^{0}$ \\[1ex]
   \hline
  \end{tabular}
\label{1}
 \end{table} 
 
 From  Eq.(16), we observe that the CP asymmetry  factor depends strongly on parameters $\theta_{13}$, $\alpha$, $\cos(\phi_{1}-\phi_{3})$ and $\cos2(2\alpha+\delta)$. In an attempt to observe the signatures of baryogenesis at neutrino oscillation experiments, we have  imposed constraints on $(\phi_{1}-\phi_{2})$ , $\alpha$, $\delta$ and $(2\alpha+\delta)$ ( dirac and majorana phases) parameters. In Fig:8  baryon asymmetry $Y_{B}$ is plotted as a function of $\alpha$ and from this plot we can observe that the disfavoured range of $\alpha$ is $100^{o}- 170^{o}$ and $280^{o}-340^{o}$. From Fig:6 we observe that the allowed range of dirac phase $\delta$ lies in the range $114^{o}- 220^{o}$. The currently constrained value of $\delta$ (or $\delta_{CP}$) by low energy neutrino oscillation experiment lies in the above mentioned range, which motivates the search of leptogenesis signatures at low energy neutrino oscillation experiments. From Eq. 16 we can observe that the dependent phase $(2\alpha+\delta)$ contributes significantly in CP asymmetry. \\
 
 \begin{figure}[H]
\includegraphics[scale=.70]{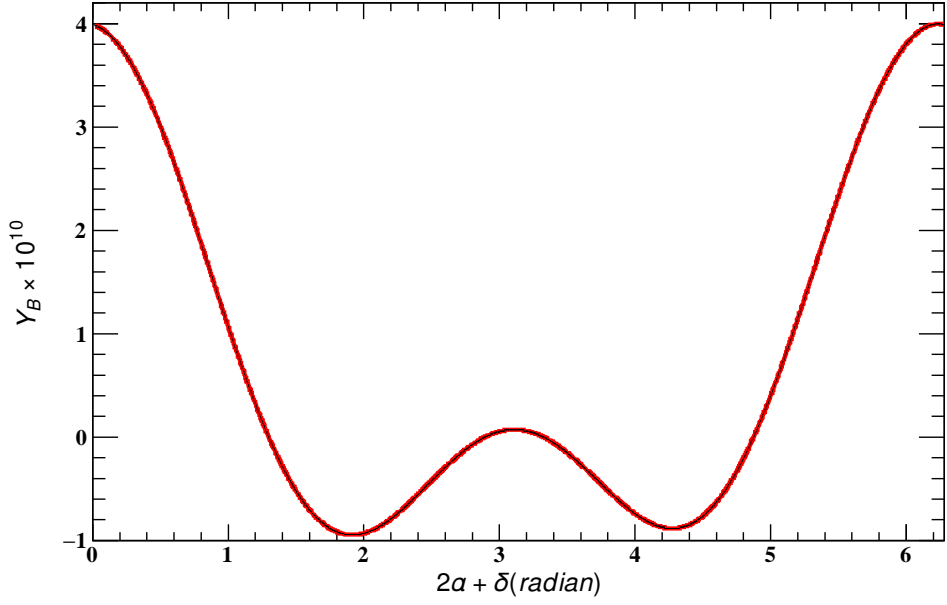}
\caption{Evolution of Baryon asymmetry parameter $Y_{B}$ as a function of the dependent phase angle $(2\alpha+\delta)$. }
\label{fig:4}      
\end{figure}

 Fig:7 illustrates, that the baryon asymmetry (originating from CP asymmetry) depends on $(2\alpha+\delta)$, a low energy dependent phase. The allowed values of this dependent phase to generate observed baryon asymmetry are $0^{o}- 57^{o}$ and $292^{o}- 360^{o}$.

 \begin{figure}[H]

\includegraphics[scale=.70]{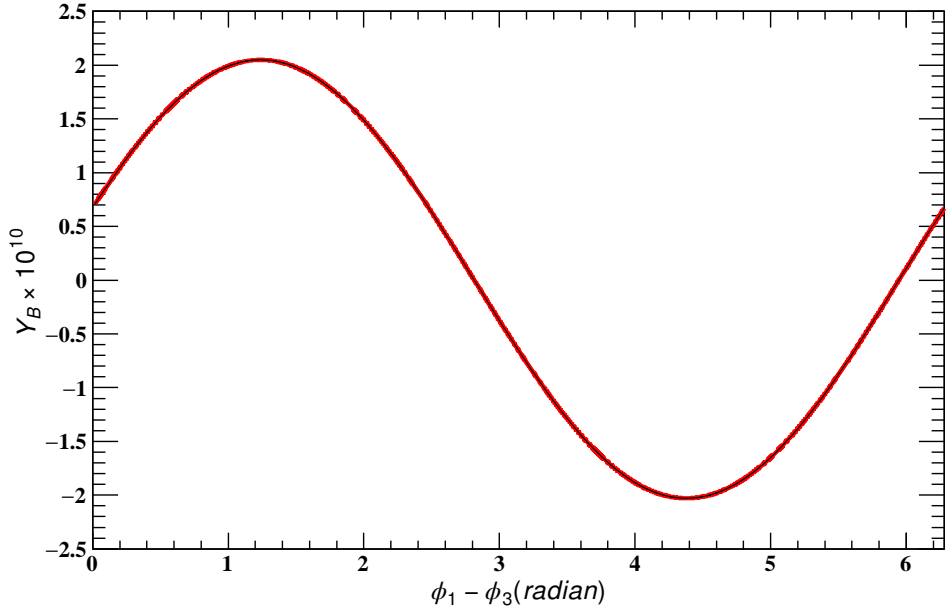}

\caption{ Evolution of Baryon asymmetry parameter $Y_{B}$ as a function of the dependent phase angle $(\phi_{1}-\phi_{3})$ .}

\end{figure}

 Fig:8. shows the dependance of baryogenesis on phases $\phi_{1}$ and $\phi_{3}$, which are the phases used for the digonalization of the right handed mass matrix as shown in Eq. (11). The allowed range of ($\phi_{1}$ - $\phi_{3}$) is  $30^{o}-110^{o}$.\\

\section{ Conclusions:}
In this work,  we have performed a study of thermal leptogenesis which is considered as mechanism responsible for the generation of baryon asymmetry. The analytical expression for $\epsilon_{1}$ (CP asymmetry) at low temperature is derived in terms of small expansion parameter $\gamma$. 
This expression is derived for non zero or the present value of oscillation angle, $\theta_{13}$ .
In an attempt to generate baryon asymmetry from CP asymmetry Boltzmann equation for $N_{N_{1}}$ (excess right handed neutrino) and $N_{B-L}$ are solved.
Fig:1 illustrates that the pattern for the production of the right handed neutrino at $K = 100$ overlaps the pattern produced for the same by $z_{eq}$, after $z = 1$. Hence we can say that to produce baryon or lepton asymmetry the value of $K$ should be less 100 or $K\textless 1$ will be preferred. This preference for the value of $K$ is observed by green line ($K = 10^{-2}$) of Fig:1.
The Fig:2 also indicates that $K \textless 1$ is preferred for the generation of $N_{B-L}$. The final efficiency expression used for the generation of the baryon asymmetry is sum of positive contribution $k_{f}^{+}(K)$ when $K\textgreater\textgreater1$ and negative contribution $k_{f}^{-}(K)$ when $K\textless\textless1$ in which both scattering and decay processes are considered.
The observed baryon asymmetry $\eta_{B} \approx 10^{-10}$ sets a lower bound limit on $\epsilon_{1}$ and therefore on $M_{1}$. For thermally genereted right handed neutrinos we get $\eta_{B} \leq 10^{-2} \epsilon_{1}$ and $M_{1} \textgreater 10^{8}$ GeV. From Fig:3 the lower bound on right handed neutrino mass $M_{1}$ in thermal initial abundance case is $\sim 5 \times 10^{8}$ GeV and in zero initial abundance case is $\sim 2 \times 10^{9}$.\\
The results illustrated in Fig:4,5,6,7 and 8 show that with the present value of $\theta_{13}$ and present bounds imposed on $\delta$ by NoVA and T2K \cite{T2K}, the neutrino experiments can be considered as one of potential source to compute the baryon asymmetry. However, a few relevant parameters are presently
unknown. Meanwhile, one can try getting interesting constraints by imposing few assumptions on the high-energy parameters (the most relevant one being that right-handed neutrinos are hierarchical)\cite{WBPD}. It would be of great interest to perform  similar calculations by including thermal corrections to CP asymmetry . This will be done in further work.
\

\begin{acknowledgement}
          This work is partially supported by Department of Physics, Lucknow University, Lucknow. I thank Dr. Jyotsna Singh for her  valuable support and guidance in completing this work from Lucknow University. 
\end{acknowledgement}\vspace{-10mm}



\end{document}